\documentclass{easychair}

\begin{document}

%
\title{Internet of Drones (IoD): Threats, Vulnerability, and Security Perspectives}

%
\titlerunning{Internet of Drones}

\volumeinfo
{The 3rd International Symposium on Mobile Internet Security (MobiSec'18), Auguest 29-September 1, 2018, Cebu, Philippines}
{37}	
{1} 

%
\author{\\
Gaurav Choudhary~, Vishal Sharma~, Takshi Gupta~, Jiyoon Kim and Ilsun You\thanks{Corresponding author: Department of Information Security Engineering, Soonchunhyang University, Asan-si-31538, South Korea.
}~\\
Department of Information Security Engineering, Soonchunhyang University, Asan-si-31538, South Korea\\
gauravchoudhary7777@gmail.com, vishal\_sharma2012@hotmail.com,\\takshi\_gupta2012@hotmail.com,74jykim@gmail.com, ilsunu@gmail.com
}

%
\authorrunning{Choudhary et al.}

\maketitle

%
\begin{abstract}
\noindent
The development of the Internet of Drones (IoD) becomes vital because of a proliferation of drone-based civilian or military applications. The IoD based technological revolution upgrades the current Internet environment into a more pervasive and ubiquitous world. IoD is capable of enhancing the state-of-the-art for drones while leveraging services from the existing cellular networks. Irrespective to a vast domain and range of applications, IoD is vulnerable to malicious attacks over open-air radio space. Due to increasing threats and attacks, there has been a lot of attention on deploying security measures for IoD networks. In this paper, critical threats and vulnerabilities of IoD are presented. Moreover, taxonomy is created to classify attacks based on the threats and vulnerabilities associated with the networking of drone and their incorporation in the existing cellular setups. In addition, this article summarizes the challenges and research directions to be followed for the security of IoD.
\newline
\newline
\textbf{Keywords}:  Internet of Drones, Security, Attacks, Vulnerabilities, Threats, UAVs.
\end{abstract}
\section{Introduction}
\label{sect:introduction}
Recent development in the area of Unmanned Aerial Vehicles (UAVs) has facilitated various opportunities at an effective cost. Due to the capability of dynamic reconfigurability, fast response and ease of deployment, UAVs can be considered as a paramount solution in many areas of surveillance, trilateral services, medical, agricultural, and transportation \cite{pajares2015overview} \cite{liu2014review}. In spite of being advantageous, high mobility is a concern for the networks that require sufficient control over these aerial vehicles. It is also one of the reasons for link distortion in UAV networking. With the modernization of Internet of Things (IoT), UAVs networking has been given a new terminology, but with a similar functioning- Internet of Drones (IoD), which supports coordination of UAVs in the sky. With this, UAVs-assisted networks operate to facilitate wireless connectivity in the areas where deployment of physical infrastructure is difficult or expensive \cite{sharma2018self}.

\begin{figure}[ht!]
\centering
\includegraphics[width=300px]{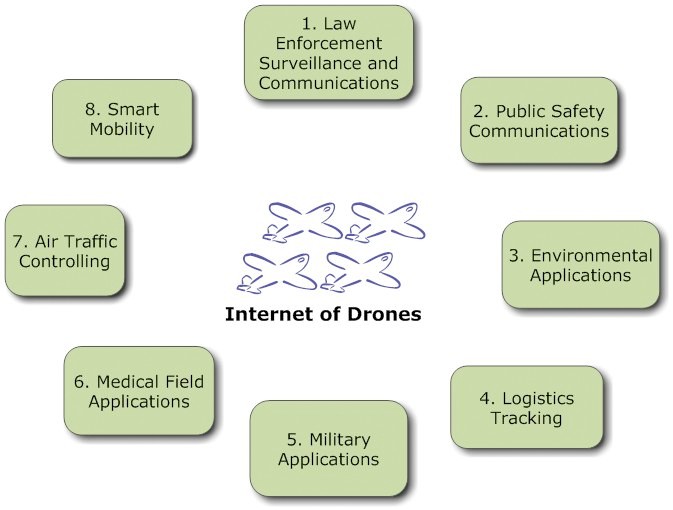}
\caption{An overview of Internet of Drones applications areas.}
\label{IoD_app}
\end{figure}
\subsection{Internet of Drones (IoD)}
\label{sect:IoD}

IoD is coined from IoT by replacing ``Things" with ``Drones" while contributing similar properties. IoD is anticipated to become an integral milestone in the development of UAVs. Gharibi et al. \cite{gharibi2016internet} defined IoD as a “layered network control architecture", which supports UAVs in coordinating. In the IoD environment, many drones combine and form a network while transmitting and receiving data from each other. IoD offers to provision for being operated remotely or through the Internet via IP addresses.
UAVs pave a way for a lot of applications, but their use-case faces a number of challenges in terms of implementation, designing, and deployment. The architectural design of drone communication lacks in standard and unification. Also, UAV-enabled communication networks suffer from an issue of dedicated spectrum sharing. UAV deployment and path planning are other issues to be considered during spectrum allocations, as these cause a high impact on energy efficiency. Moreover, UAV communications require consistency with large and reliable networks along with their security. IoD can be considered as a significant solution to supports coordination of UAVs and other issues related to UAV communications. IoD enhances the storing capabilities and increases the facilitation of algorithmic solutions, which in turn allow solutions inspired by Machine Learning and Artificial Intelligence to be used for cooperative formations \cite{chen2017liquid}.
In terms of applications, IoD can be used as airspace allocation interface, as well as a provider of ultra-high reliability, navigation and accuracy to drones in various kind of areas as presented in Fig.\ref{IoD_app}. Various technologies like long-range wide-area network (LoRaWAN), low-power wide-area network (LPWAN), Narrow Band IoT (NB-IoT) are sustainable solutions in the deployment of IoD in amalgamation with IoT\cite{you2018enhanced}.

The adoption of IoD with the properties of Low cost, small size and high reconfigurability, operational functionality and real-time accountability is the need of the hour\cite{hall2016internet}.  The market trends are floating towards the adoption of IoD with following characteristics:
\begin{itemize}
\item  IoD supports device to device as well as device to multi-device communications.
\item  IoD facilitates connectivity to the contextual networks.
\item  IoD can be operated as a data gathering and information management service.
\item IoD helps to navigate and enhance the survivability of vehicles through V2X or V2V mode (Drone to Drone mode).
\end{itemize}

Irrespective of the advancements and plethora of solutions for drone communications, Security is still a primary concern in IoD as transmissions involve sensitive and critical information. UAVs are usually operated remotely while receiving control and command messages from ground stations and coordinated in IoD; however, some cases may involve autonomous UAVs as well. These command and control messages are transmitted over different channels and variable transmission rate, which needs considerable effort for management and control \cite{sharma2018dptr}. Hence, the security of IoD is one of the most important requirements in the true utilization of multiple UAVs.

\subsection{IoD: Application Scenario}
\label{sect:IoD_app}
Rapidly deployable and reliable mission-critical communication applications are the basic requirements for a successful mission. IoD plays a significant role in such type of applications. IoD can facilitate various services for the military applications and surveillance. IoD devices/UAVs work as an intermediate platform for various sensor and devices which has been deployed to collect data and information in mission-oriented tasks.

\begin{figure}[ht!]
\centering
\includegraphics[width=450px]{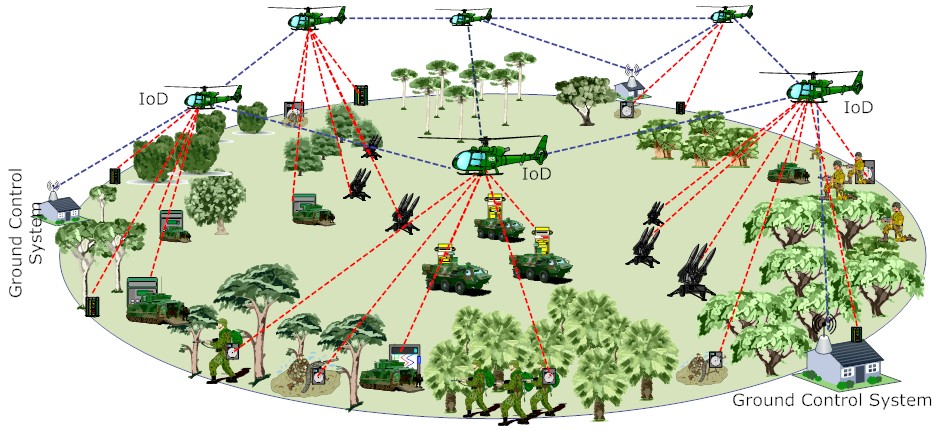}
\caption{An exemplary illustration of IoD-assisted battlefield scenario.}
\label{IoD_sci}
\end{figure}

One of the most suitable application scenarios of IoD, UAVs-enabled battlefield \cite{springer2013military}, is described in fig \ref{IoD_sci}. In this application scenario, IoD can be used as an airspace allocation interface which helps to navigate the neighboring military UAVs. IoD collects the data from the deployed sensors along with specific types of military equipment. Considering the domain of application and content, the collected information in such a scenario is very critical and should be protected from intruders.

\subsection{IoD as a Service}
\label{sect:IoD_Next}
IoD is an integral part of the future Internet, which may be used for next-generation applications like smart tracking, smart parking, air traffic control, and cellular networks. The recent development in the communications network and drone applications has attracted the researchers and market for the advancement of IoD. The communication incorporating IoD in numerous applications is another footprint for the next-generation development of these networks. IoD has characteristics similar to smaller fleets with limited batteries that send and receive a large amount of data \cite{zancul2016business}\cite{andersen2017enabling}.

IoD can be used to allow users to develop the connected devices workflow via combining various integrated drones and databases with a proper resource planning \cite{sharma2017uav}. For these mechanisms, various Machine Learning and data mining techniques can be used for processing diversified data\cite{abomhara2015cyber}.

The next generation IoD is entitled to be integrated with lots of third parties businesses and object-oriented service over the sky. The future IoD market incorporates with the variety of tasks in the collaboration mode. UAVs in IoD have abilities for multitasking and operate in urban and rural environments. In addition, this next-generation IoD should cope with security solution to mitigate the risks associated with their missions.

\section{Security threats and vulnerability issues with IoD}
\label{sect:IoD_threats}
The threats are considered as critical factors because these are responsible for exploiting vulnerabilities by introducing unwanted alterations \cite{kanellos2016safety}. This paper addresses various vulnerabilities and threats associated with IoD as shown in Table \ref{table_threat}. These threats are considered similar to a general IoD system as most of the data with these systems are very sensitive.

The sensitive pieces of information are gathered or processed through IoD while classifying them with their respective tasks. The leakage of such information can lead to a big loss in terms of privacy and trust \cite{akram2017security}.

\begin{table}[!ht]
\fontsize{8}{10}\selectfont
\centering
\caption{Threats and vulnerabilities associated with the Internet of Drones (IoD).}
\label{table_threat}
\begin{tabular}{|l|l|l|l|}
\hline
\multicolumn{1}{|c|}{\textbf{Threat and vulnerability of IoD}}                                                                                                & \multicolumn{1}{c|}{\textbf{\begin{tabular}[c]{@{}c@{}}Affected security \\ Parameter\end{tabular}}} & \multicolumn{1}{c|}{\textbf{Impact}}                                                                                                                                               & \multicolumn{1}{c|}{\textbf{\begin{tabular}[c]{@{}c@{}}Target \\ Components\end{tabular}}} \\ \hline
\begin{tabular}[c]{@{}l@{}}An attacker can compromise \\ keys and capture the \\ communications\end{tabular}                                                  & \begin{tabular}[c]{@{}l@{}}Integrity and \\ confidentiality\end{tabular}                             & The breach of privacy                                                                                                                                                              & \begin{tabular}[c]{@{}l@{}}Networks, \\ platforms\end{tabular}                    \\ \hline
\begin{tabular}[c]{@{}l@{}}An attacker can modify or \\ fabricate information in the\\ transmission\end{tabular}                                              & \begin{tabular}[c]{@{}l@{}}Integrity and \\ confidentiality\end{tabular}                             & \begin{tabular}[c]{@{}l@{}}The user can misguide\\ and obtained falsified data\end{tabular}                                                                                        & \begin{tabular}[c]{@{}l@{}}  Networks, \\ platforms\end{tabular}                    \\ \hline
\begin{tabular}[c]{@{}l@{}}An attacker can change \\ authorization or access \\ permissions by gaining\\ access controls\end{tabular}                         & \begin{tabular}[c]{@{}l@{}}Confidentiality \\ and availability\end{tabular}                          & \begin{tabular}[c]{@{}l@{}}The IoD device can be\\ captured by non-legitimate\\ users. The permission can \\ be altered which leads to \\ conflicts and system crash.\end{tabular} & \begin{tabular}[c]{@{}l@{}} Networks, \\ platforms, \\ services\end{tabular}       \\ \hline
\begin{tabular}[c]{@{}l@{}}An attacker can perform \\ physical attacks as well as \\ DoS attacks by excessive \\ false requests\end{tabular}                  & Availability                                                                                         & \begin{tabular}[c]{@{}l@{}}The attacker aims to\\ destroy the IoD device \\ which is used for \\ certain applications.\end{tabular}                                                & \begin{tabular}[c]{@{}l@{}} Networks, \\ platforms, \\ services\end{tabular}       \\ \hline
\begin{tabular}[c]{@{}l@{}}The coordination of IoD \\ can be altered by hacking \\ or GPS spoofing\end{tabular}                                               & Availability                                                                                         & \begin{tabular}[c]{@{}l@{}}The attacker capture the\\ IoD devices or altered the\\ coordination which may \\ lead to collisions.\end{tabular}                                      & \begin{tabular}[c]{@{}l@{}} Networks, \\ platforms, \\ services\end{tabular}       \\ \hline
\begin{tabular}[c]{@{}l@{}}The attacker can affect \\ the performance of IoD \\ by increasing\\ resource depletion rate \\ (fuel, battery etc.)\end{tabular} & Availability                                                                                         & \begin{tabular}[c]{@{}l@{}}The resource depletion rate \\ reduces the performance of \\ IoD in the target missions, \\ which may lead to mission \\ failure.\end{tabular}         & \begin{tabular}[c]{@{}l@{}} Platforms, \\ services\end{tabular}                    \\ \hline
\begin{tabular}[c]{@{}l@{}}The attacker can capture \\ the IoD and turn off the \\ services like camera, \\ Internet facility etc.\end{tabular}               & \begin{tabular}[c]{@{}l@{}}Confidentiality \\ and availability\end{tabular}                          & \begin{tabular}[c]{@{}l@{}}The IoD are assigned for\\ a dedicated task, switching\\ off the components and\\ service leads to failures \\ in their dedicated jobs.\end{tabular}   & \begin{tabular}[c]{@{}l@{}}Networks, \\ platforms, \\ services\end{tabular}       \\ \hline
\end{tabular}
\end{table}

IoD plays a significant role in the network communication in case of Public Safety Communications (PSCs) or emergencies. In such case, the attacker or an intruder can compromise the keys and captures the communication which leads to the breach of privacy. The attacker can target the vulnerability within the IoD and its application platforms to gain keys. The access to this information can be fabricated or modified by the intruder which may misguide the receivers. The access control of IoD is a crucial parameter; therefore, security concerns about access and authorization should be emphasized. If an intruder alters the access control information, then the IoD configuration can be changed easily. This change in configuration creates lots of conflicts with respective IoD. The services-oriented IoD is used for dedicated jobs like real-time surveillance, air traffic controller etc. In such cases, an attacker may capture a node and turn off the services assigned to entire IoD, which will lead to multiple mission failures.

The critical missions oriented IoD can be affected by Denial-of-Service (DoS) \cite{peng2007survey} or Distributed Denial-of-Service (DDoS) \cite{douligeris2004ddos} attacks by sending excessive requests. This will affect the availabilities of IoD in a critical mission. For an IoD, the resources are limited; therefore, an attacker can target to exploit excessive resource utilization. For this, an attacker may enable excessive services which may lead to a high depletion rate.

Many more threats can be considered which affects the availability of IoD. The waypoint alteration, Global Positioning System (GPS) spoofing, IoD Freezing are considerable attacks against the availability. The attacker can change coordination points with these attacks which can lead to a collision in IoD. In such type of threats and vulnerabilities, additional security and countermeasures are needed to handle vulnerability exploitations. The security solutions like Intrusion Detection Systems (IDS) \cite{liu2011research}, cryptographic mechanisms, and security policies can help to protect IoD \cite{sharma2018three}.

\section{Attacks in IoD}
\label{sect:IoD_Attacks}
\begin{figure}[ht!]
\centering
\includegraphics[width=500px]{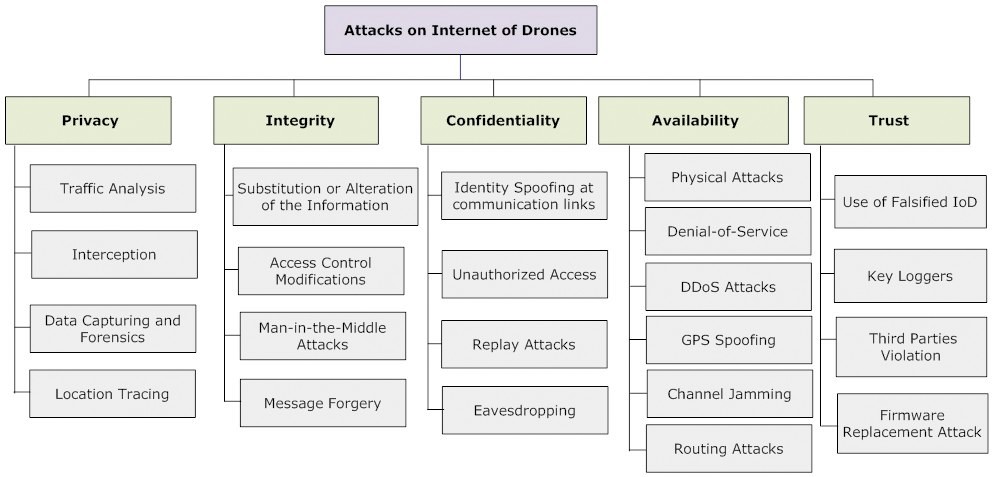}
\caption{A taxonomy of different types of known attacks in IoD.}
\label{IoD_sc}
\end{figure}

The primary victims of compromising the security of IoD are various system components like IoD devices, networks, and communication links \cite{yampolskiy2013taxonomy}. The threats, as discussed in the Section \ref{sect:IoD_threats}, are responsible for the security breach and lead to a big loss in terms of resources, trust, and availability \cite{javaid2012cyber} \cite{mansfield2013unmanned}. On the basis of these vulnerabilities, the attacks can be classified into five main domains as presented in fig \ref{IoD_sc}.

\begin{enumerate}
\item
\textbf{Privacy}:  Privacy is an important concern for data-oriented security in IoD. The data is collected and processed through IoD and processing of data increase the possibilities of threats and vulnerabilities \cite{pauner2015drones}. The attacker targets IoD to gain sensitive information through various approaches. The following attacks affect the privacy of IoD.
 \begin{itemize}
 \item \textbf{Traffic analysis}: The traffic analysis is performed to examine the IoD traffic to get some useful information from the IoD devices and networks.  The traffic contains packets shared between IoD and ground control systems. The forensics of packets in the traffic reveals sensitive information. The packets include information like location, connected IoD with sensors, and captured data from sensors.
 \end{itemize}

  \begin{itemize}
  \item \textbf{Interception}: In the interception, intruder involves someone who routinely monitors network traffic. It is very difficult to find such an intruder who is passively monitoring the network. In critical missions, IoD contains sensitive information; therefore, tracking and monitoring of IoD can be dangerous for the agencies that are responsible for such missions.
   \end{itemize}
 \begin{itemize}
   \item \textbf{Data capturing and forensics}: Through traffic analysis, a huge amount of data can be collected from IoD. Even if the encrypted data do not reveal any useful information, data forensics help to get sensitive information from collected data. Thus, it is desirable to formulate solutions for preventing information breach in case the forensics-based mechanisms are applied for attacking IoD.
 \end{itemize}

  \item
   \textbf{Integrity}: The integrity defines that data in IoD should be consistent, accurate, and trusted. The transmission should not be altered in the communication by the non-legitimate users or attackers \cite{hartmann2013vulnerability}. Some of the commonly used mechanisms for protection of data integrity are hash functions, checksums etc. The integrity of IoD is affected by the following attacks:

    \begin{itemize}
    \item \textbf{Substitution or alteration of the information}: The alteration is the concept of adding false or wrong information in the communications and change the original meaning of data. Various forms of alterations include modification, fabrication, substitutions, and data injections which modifies the data used in the communication of IoD. The alteration of data misguides the users with fabricated information.

     \item \textbf{Access-control modification}: The access controls are the rules and policies that govern how other devices in IoD communicate and how a user accesses the data. The access control is the mind of a body which provides instructions to the IoD. If an attacker gains access controls, then the attacker can change all the permissions, privileges and authorizations, which may lead to huge losses.

    \item \textbf{Man-in-the Middle attacks}: Man-in-the-Middle attacks permits attackers to capture the data on the communication between IoD and sensors. The Rogue Access Point is used to own wireless access point and trick nearby devices to join its domain in the IoD communications. Through these access points, network traffic can be manipulated by the attacker \cite{kamthan2017uavs}. There are various solutions against the prevention of Man-in-the-Middle attacks, which include Strong Wired Equivalent Privacy (WEP)/Wi-Fi Protected Access (WAP) encryption on the access points, Hyper Text Transfer Protocol Secure (HTTPS), and Public Key-based authentications.

    \item \textbf{Message forgery}:  Under the message forging attack in IoD, login request message of previous sessions over the public/open channel are forged during authentication protocol execution. After that, an attacker can modify and retransmit the message to the user.
      \end{itemize}
  \item  \textbf{Confidentiality}: The confidentiality ensures that the information cannot be leaked to the non-legitimate users. Many attacks against IoD and ground control station are a result of deficiencies in the security \cite{akram2017security}. The confidentiality is affected by non-legitimate users’ access to IoD and that retrieve useful information.
\begin{itemize}
\item \textbf{Identity spoofing}: In the identity spoofing, the attacker successfully masquerades as a legitimate user in the IoD network with the spoofing ID of the legitimate user and gain access to the IoD network and communication links. Encrypted IDs or one-time usable pseudo IDs can be an efficient solution against the prevention of such attacks.

\item \textbf{Unauthorized access}: Unauthorized access is when someone gains access to the IoD server and services using someone else's account or other methods like duplicate IDs. This attack leads to a risk of unauthorized discloser of critical information from the IoD.

\item \textbf{Replay attacks}: In the replay attacks, a malicious user sniffs IoD network and bypasses the security mechanisms by replaying requests in the IoD server. The replay attack can be performed in various ways. In order to prevent such an attack, the authentication mechanisms should use fresh message requests in a secured manner in the IoD networks for obtaining data and start communications.

\item \textbf{Eavesdropping}: The eavesdropping can be defined as unauthorized real-time interception of IoD communications. The eavesdropping is a potentially dangerous attack because it allows an attacker to retrieve confidential information exchanged between devices in IoD.  Lack of authentication and unencrypted data in the communication lead to such attacks.
\end{itemize}

 \item
  \textbf{Availability}:  Availability is defined as the services initiated immediately when needed for maintaining a correct functioning. Availability of information is to ensure that legitimate users are able to access the information on the basis of their requirements. IoD is operated in mission-oriented fields or areas, therefore, the availability of IoD is major concerns in respect of security \cite{yampolskiy2013taxonomy}. The availability of IoD can be affected by following attacks:
  \begin{itemize}
  \item \textbf{Physical attacks}: These types of attacks are performed on the hardware components. These attacks have the prime motivation for destroying the devices. IoD devices are expensive; therefore the protection against physical attacks is a considerable issue.

  \item \textbf{DoS and DDoS}:
DoS is defined as denying accessibility of resources or preventing the legitimate users from accessing service from the designated resources. IoD needs a communication channel to send and receive data \cite{rodday2016exploring}. If the attacker performs flooding requests on these channels, the network is interrupted, which leads to resource unavailability.

  \item \textbf{GPS spoofing}: GPS is used to determine the position of the vehicle and provide waypoints to fly at the designated target. The attacker can modify the content of the received GPS signals or generate the spoofing signals with the help of GPS signal generators \cite{giray2013anatomy}. The delays in GPS signals will also cause a big loss in IoD as this can break the coordination resulting in mid-air collisions. Some of the countermeasures that can be considered against the GPS spoofing are:  IoD should use authenticated and encrypted signals, and timely check the signal strength of GPS with relative and absolute values, and check the receiving time intervals.

 \item \textbf{Channel jamming}: The preliminary objective of jamming is to intentionally disrupt the communication channel of IoD \cite{bhattacharya2010game}. The vulnerabilities in the communication channel can affect the IoD. Jamming can cause the collision of IoD or unavailability of services \cite{rudinskas2009security}.

\item \textbf{Routing attacks}: The routing attacks include flooding, node isolation, location discloser attacks, etc\cite{yampolskiy2013taxonomy}.
 \end{itemize}

  \item
   \textbf{Trust}: Various new risk is associated with the new stage of technologies. Trust is an important asset and influenced by many measurable and non-measurable properties like goodness, strength, reliability, availability, and ability. The security and privacy are two faces of trust. The trust relationships rely on IoD developing and deploying organizations~\cite{oleson2011trust}. Various threats to trust are obtained due to misconfiguration and limited incorporation of security mechanisms in the development and deployment phases.

 \begin{itemize}
   \item \textbf{Use of falsified IoD}: In the deployment phase of IoD, the fabricated IoD can be replaced with the legitimate IoD devices. The fabricated IoD sends data to the eavesdropper who has access to these IoD. In the critical mission, such type of IoD can create a big trouble and sensitive information can be leaked.

    \item \textbf{Key loggers}: keyloggers are considered as an internal threat. In the development and deployment of IoD, keyloggers are incorporated with the software of IoD. The sensitive information is forwarded by the attached keyloggers to the attacker.

      \item \textbf{Third party violations}: Some applications involved trusted third parties in the IoD communications like certifications and central management authorities. If trusted parties sale or disclose data obtained in the applications or deployment, this may lead to a big loss in terms of finance or intellectual properties.

      \item \textbf{Firmware replacement attack}: When any IoD device is sent for repair or maintenance phase, the firmware or software are updated with various new updates and features. If an attacker exploits such a firmware upgrade by replacing the object with malicious software or vulnerabilities, entire IoD sessions can be captured by the attackers during the mission.

 \end{itemize}
\end{enumerate}
\section{Challenges and future research}
\label{sect:IoD_Challenges}
IoD is exponentially emerging technology facilitating connectivity to the Internet.  This emerging technology is facing lots of security threats and issues. Therefore, the objective of IoD security is the enhancement of protection in terms of security for IoD and its infrastructure. Some future developments are required to improve the IoD for further applications. At a glance, the IoD is facing issues in the development, deployment, communication, and coordination phases. Some of the most significant challenges and research to target include \begin{enumerate}
\item  \textbf{Low overheads and dynamic load balancing}: The overheads of IoD in memory, energy, and delays lead to various issues like resource degradations and low efficiency. Dynamic load balancing in IoD empowers the operational efficiency. The real-time capturing and processing of data requires a lot of energy and memory. Therefore, supporting various tasks with low overheads is an open challenge. Dynamic load balancing in IoD can be considered as an effective solution and a topic to follow for further research.
\item \textbf{Survivability and lifetime}: IoD involves UAVs that are battery operated and have limited power; therefore, the coordination should be facilitated at low power. The survivability is affected by the falsified signals and starvation conditions. These problems increase the depletion rate of resources like memory, energy etc. Improving IoD survivability, and enhancing the lifetime with the limited resources requires a considerable attention from the research community\cite{sharma2018secure}\cite{long2018energy}. It should be compulsory to accommodate better survivability that requires new and innovative technologies, with effective and safer capabilities in the automation and optimization of mission planning in unstructured environments of IoD.

\item \textbf{Low cost of operation and high throughput}: The cost is also a considerable factor in the development and deployment of IoD. The manufacturing cost is directly proportional to the requirements of IoD. Non-verified deployment of IoD leads to huge Capital Expenditures (CAPEX) / Operating Expenses (OPEX) issues which affect the performance in terms of cost. In addition, using IoD for enhancing the services to larger groups of isolated users is entirely dependent on the cost of operation. Thus, low-cost solutions should be developed for enhancing the utilities of IoD. As suggested in the existing studies, throughput is considerably affected through the cost of operation, and there exists a tradeoff between the cost of operation and average throughput attained in IoT, which has to be balanced for effective communications \cite{sharma2017quat}.

\item  \textbf{Performance and reliability}: The performance of IoD is accomplished with the mission completion time and efficient resource utilization.  Performance and reliability depend on high-quality manufacturing by considering the necessary prerequisites while delivering end solutions for IoD. To understand resource efficiency and reliability, consumption-focused indicators should be incorporated in IoD devices. The monitoring of such resources is evaluated in terms of performance. The efficient use of resources leads to successful missions \cite{li2017optimal}. Therefore, the resource utilization is considered as a significant parameter for performance and reliability in IoD.

\item  \textbf{Dynamic topologies and adaptive routing mechanisms} The dynamic topologies and adaptive routing mechanisms facilitate on-demand and realistic environments for IoD. The dynamic routing topologies support waypoint coordination in IoD networks \cite{sharma2018offrp}. The delays exhibited through the conventional routing algorithms do not efficiently accommodate IoD coordination. Therefore, adaptive routing mechanisms should be considered for dynamically changing topologies in IoD \cite{wu2018joint}.

\item  \textbf{Low failure rate of IoD}: IoD has been designed, developed, and used for specific applications like environmental monitoring, agriculture, etc. To support these services, IoD formations should be considered with resource constraints, dynamic topology, adaptability, scalability, security, and QoS support to the users. The failure rate of the network degrades the overall performance of IoD. Therefore, the failure rate of the network should be minimized to improve the overall performance of IoD \cite{sharma2015self}.

 \item  \textbf{Emphasized security solutions for IoD}: The recent development in IoD emphasizes the data rate and security. The security is a combined aspect of confidentiality, integrity, authentication, and non-repudiation of transmitted data in IoD networks \cite{haque2018new}. The eavesdropping, network jamming, weak authentication, and mobility management are major open issues in IoD communications.

\item  \textbf{Efficient deployment of countermeasures and maintenance}: The security solution incorporates an Intrusion Detection System, authentication mechanisms, and cryptographic algorithms. Traditional mechanisms are slow in speed, large in size and consume more power and may fail to provide necessary protection for data in the network \cite{kamthan2017uavs}. Therefore, the efficient deployment of countermeasures and maintenance can save the additional computational cost and can prevent excessive power consumptions.

\item  \textbf{End-to-end connectivity and cooperation}:  The end to end connectivity reflects the coordination efficiency. In the IoD system, connectivity between the source and the destination should always be retained. The end to end connectivity enhances the cooperation among neighboring IoD and other sensors. Some application areas, like real-time monitoring, require an extensive end to end connectivity for better coverage and cooperation \cite{orsino2017effects}.
\end{enumerate}

Securing IoD is not an easy task because of the difference in communication standards and range of applicability. IoD nodes are prone to various types of attack in a network such as Sybil attack, Wormhole attack, Sinkhole attack, or Impersonation attack, Coagulation attack\cite{sharmacoagulation}. Therefore, timely strategies, counter-mechanisms, and easily updatable solutions are required to counterfeit the harmful effects of various kinds of known and unknown cyber attacks.

\section{Conclusions}
\label{sect:Conclusions}
IoD is the emerging technology of uniting drones, which consist of analyzing the continuously evolving data from heterogeneous sources for creating a new era of real-life applications connected to the Internet. Attacks on IoD present a serious problem to the actual operational use of networked UAVs. Security threats and vulnerabilities can lead to an attack on confidentiality, integrity, authenticity, and availability of IoD. Message security and control signal protections are achieved by cryptographic mechanisms. However, security issues like unauthorized access, malicious control, illegal connection, or other malicious attacks need strategic solutions without adverse effects on the performance. Identification of threats and their mitigation in IoD pose multiple research issues that are to be taken care of through secure and efficient approaches.

In future, the work will be presented on the threat mitigation mechanisms and vulnerability assessments methods for IoD.
\section*{Acknowledgement}
This work was supported by 'The Cross-Ministry Giga KOREA Project' grant funded by the Korea government(MSIT) (No.GK18N0600, Development of 20Gbps P2MP wireless backhaul for 5G convergence service)

%
\label{sect:bib}
\bibliographystyle{abbrv}
\bibliography{ref-mobisec18}
------------------------------------------------------------------------------
\section*{Author Biography}
\vspace*{1em}
\begin{biography}{Gaurav Choudhary}{gau} received the B.Tech. degree in Computer Science and Engineering from Rajasthan Technical University in 2014 and the Master Degree in Cyber Security from Sardar Patel University of Police in 2017. He is currently pursuing Ph.D. degree in the Department of Information Security Engineering, Soonchunhyang University, Asan, South Korea. His areas of research and interests are UAVs, Mobile and Internet security, IoT security, Network security, and Cryptography.
\end{biography}
\vspace*{1em}
\begin{biography}{Vishal Sharma}{a3} received the Ph.D. and B.Tech. degrees in computer science and engineering from Thapar University (2016) and Punjab Technical University (2012), respectively. He worked at Thapar University as a Lecturer from Apr’16-Oct’16. From Nov. 2016 to Sept. 2017, he was a joint post-doctoral researcher in MobiSec Lab. at Department of Information Security Engineering, Soonchunhyang University, and Soongsil University, Republic of Korea. Dr. Sharma is now a Research Assistant Professor in the Department of Information Security Engineering, Soonchunhyang University, The Republic of Korea. Dr. Sharma received three best paper awards from the IEEE International Conference on Communication, Management and Information Technology (ICCMIT), Warsaw, Poland in April 2017; from CISC-S’17 South Korea in June 2017; and from IoTaas Taiwan in September 2017. He is the member of IEEE, a professional member of ACM and past Chair for ACM Student Chapter- TU Patiala. He has authored/coauthored more than 60 journal/conference articles and bookchapters. He serves as the program committee member for the Journal of Wireless Mobile Networks, Ubiquitous Computing, and Dependable Applications (JoWUA). He was the track chair of MobiSec’16 and AIMS-FSS’16, and PC member and reviewer of MIST’16 and MIST’17, respectively. He was the TPC member of ITNACIEEE TCBD’17 and serving as TPC member of ICCMIT’18, CoCoNet’18 and ITNAC-IEEE TCBD’18. Also, he serves as a reviewer for various IEEE Transactions and other journals. His areas of research and interests are 5G networks, UAVs, estimation theory, and artificial intelligence.
\end{biography}
\vspace*{0.5em}
\begin{biography}{Takshi Gupta}{a1} received the B.Tech. degree in Computer Science and Engineering
from Punjab Technical University in 2012 and the PG Diploma in Business Administration from SCDL, Symbiosis University, India in 2014 both with distinction. She is currently associated with the MobiSec Lab, Department of Information Security Engineering, Soonchunhyang University, Asan, South Korea. Prior to this, she worked at Kochar InfoTech followed by Gen-XT Infosystems and was a co-founder at Future Info-Systems. Her areas of research and interests are database management and security, artificial intelligence, semantic webs, and human resource management.
\end{biography}

\vspace*{0.5em}
\begin{biography}{Jiyoon Kim}{a5} received the B.S. degree in information security engineering from Soonchunhyang University, Asan, South Korea, where he is currently pursuing the master degree in the Department of Information Security Engineering. His current research interests include mobile Internet security, IoT security, and formal security analysis.
\end{biography}

\vspace*{3.5em}
\begin{biography}{Ilsun You}{a4} received the M.S. and Ph.D. degrees in computer science from Dankook University, Seoul, Korea, in 1997 and 2002, respectively. He received the second Ph.D. degree from Kyushu University, Japan, in 2012. From 1997 to 2004, he was at the THINmultimedia Inc., Internet Security Co., Ltd. and Hanjo Engineering Co., Ltd. as a research engineer. Now, he is an associate professor at Department of Information Security Engineering, Soonchunhyang University. He has served or is currently serving as a main organizer of international conferences and workshops such as MobiWorld, MIST, SeCIHD, AsiaARES, and so forth. Dr. You is the EiC of Journal of Wireless Mobile Networks, Ubiquitous Computing, and Dependable Applications (JoWUA). He is in the Editorial Board for Information Sciences (INS), Journal of Network and Computer Applications (JNCA), International Journal of Ad Hoc and Ubiquitous Computing (IJAHUC), Computing and Informatics (CAI), Journal of High Speed Networks (JHSN), Intelligent Automation \& Soft Computing (AutoSoft), and Security and Communication Networks (SCN). His main research interests include Internet security, authentication, access control, and formal security analysis. He is a Fellow of the IET and a senior member of the IEEE.
\end{biography}
\vspace*{0.5em}

\end{document}